\pdfoutput=1

\documentclass[12pt,a4paper]{article}

\usepackage{ifthen} 
\newboolean{pdflatex}
\setboolean{pdflatex}{true} 

\newboolean{articletitles}
\setboolean{articletitles}{true} 

\newboolean{uprightparticles}
\setboolean{uprightparticles}{false} 

\newboolean{inbibliography}
\setboolean{inbibliography}{false} 

\usepackage{microtype}
\usepackage{lineno}  
\usepackage{xspace} 
\usepackage{caption} 

\usepackage{graphicx}  
\usepackage{color}
\usepackage{colortbl}
\graphicspath{{./figs/}} 

\usepackage{amsmath} 
\usepackage{amssymb}
\usepackage{amsfonts}
\usepackage{upgreek} 

\newcommand*\patchAmsMathEnvironmentForLineno[1]{%
\expandafter\let\csname old#1\expandafter\endcsname\csname #1\endcsname
\expandafter\let\csname oldend#1\expandafter\endcsname\csname
end#1\endcsname
 \renewenvironment{#1}%
   {\linenomath\csname old#1\endcsname}%
   {\csname oldend#1\endcsname\endlinenomath}%
}
\newcommand*\patchBothAmsMathEnvironmentsForLineno[1]{%
  \patchAmsMathEnvironmentForLineno{#1}%
  \patchAmsMathEnvironmentForLineno{#1*}%
}
\AtBeginDocument{%
\patchBothAmsMathEnvironmentsForLineno{equation}%
\patchBothAmsMathEnvironmentsForLineno{align}%
\patchBothAmsMathEnvironmentsForLineno{flalign}%
\patchBothAmsMathEnvironmentsForLineno{alignat}%
\patchBothAmsMathEnvironmentsForLineno{gather}%
\patchBothAmsMathEnvironmentsForLineno{multline}%
\patchBothAmsMathEnvironmentsForLineno{eqnarray}%
}

\usepackage{hyperref}    
\usepackage[all]{hypcap} 

\def\lhcb {\mbox{LHCb}\xspace}

\def\MagUp {\mbox{\em Mag\kern -0.05em Up}\xspace}

\ifthenelse{\boolean{uprightparticles}}%
{

 \def\Pmu         {\ensuremath{\upmu}\xspace}                 
 \def\Pnu         {\ensuremath{\upnu}\xspace}                 
                  
 \def\Ppi         {\ensuremath{\uppi}\xspace}                 
                  
 \def\Prho        {\ensuremath{\uprho}\xspace}                 
                  
 \def\Ptau        {\ensuremath{\uptau}\xspace}                 
                  
 \def\Pphi        {\ensuremath{\upphi}\xspace}

 \def\Ppsi        {\ensuremath{\uppsi}\xspace}                 
 \def\Pomega      {\ensuremath{\upomega}\xspace}                 

 \def\PDelta      {\ensuremath{\Delta}\xspace}                 
 \def\PXi      {\ensuremath{\Xi}\xspace}                 
 \def\PLambda      {\ensuremath{\Lambda}\xspace}                 
 \def\PSigma      {\ensuremath{\Sigma}\xspace}                 
 \def\POmega      {\ensuremath{\Omega}\xspace}                 
 \def\PUpsilon      {\ensuremath{\Upsilon}\xspace}                 
 
 \def\PB      {\ensuremath{\mathrm{B}}\xspace}                 
                  
 \def\PD      {\ensuremath{\mathrm{D}}\xspace}

 \def\PJ      {\ensuremath{\mathrm{J}}\xspace}                 
 \def\PK      {\ensuremath{\mathrm{K}}\xspace}

 \def\Pb      {\ensuremath{\mathrm{b}}\xspace}                 
 \def\Pc      {\ensuremath{\mathrm{c}}\xspace}

 \def\Pi      {\ensuremath{\mathrm{i}}\xspace}

 \def\Ps      {\ensuremath{\mathrm{s}}\xspace}                 
                  
 \def\Pu      {\ensuremath{\mathrm{u}}\xspace}

}
{

 \def\Pmu         {\ensuremath{\mu}\xspace}                 
 \def\Pnu         {\ensuremath{\nu}\xspace}                 
                  
 \def\Ppi         {\ensuremath{\pi}\xspace}                 
                  
 \def\Prho        {\ensuremath{\rho}\xspace}                 
                  
 \def\Ptau        {\ensuremath{\tau}\xspace}                 
                  
 \def\Pphi        {\ensuremath{\phi}\xspace}

 \def\Ppsi        {\ensuremath{\psi}\xspace}                 
 \def\Pomega      {\ensuremath{\omega}\xspace}                 
 \mathchardef\PDelta="7101
 \mathchardef\PXi="7104
 \mathchardef\PLambda="7103
 \mathchardef\PSigma="7106
 \mathchardef\POmega="710A
 \mathchardef\PUpsilon="7107
                  
 \def\PB      {\ensuremath{B}\xspace}                 
                  
 \def\PD      {\ensuremath{D}\xspace}

 \def\PJ      {\ensuremath{J}\xspace}                 
 \def\PK      {\ensuremath{K}\xspace}

 \def\Pb      {\ensuremath{b}\xspace}                 
 \def\Pc      {\ensuremath{c}\xspace}

 \def\Pi      {\ensuremath{i}\xspace}

 \def\Ps      {\ensuremath{s}\xspace}                 
                  
 \def\Pu      {\ensuremath{u}\xspace}

}

\makeatletter
\ifcase \@ptsize \relax
  \newcommand{\miniscule}{\@setfontsize\miniscule{4}{5}}
\or
  \newcommand{\miniscule}{\@setfontsize\miniscule{5}{6}}
\or
  \newcommand{\miniscule}{\@setfontsize\miniscule{5}{6}}
\fi
\makeatother

\DeclareRobustCommand{\optbar}[1]{\shortstack{{\miniscule (\rule[.5ex]{1.25em}{.18mm})}
  \\ [-.7ex] $#1$}}

\def\mup        {{\ensuremath{\Pmu^+}}\xspace}
\def\mun        {{\ensuremath{\Pmu^-}}\xspace} 

\def\tauon      {{\ensuremath{\Ptau}}\xspace}

\def\neu        {{\ensuremath{\Pnu}}\xspace}

\def\neum       {{\ensuremath{\neu_\mu}}\xspace}

\def\uquark    {{\ensuremath{\Pu}}\xspace}

\def\squark    {{\ensuremath{\Ps}}\xspace}

\def\cquark    {{\ensuremath{\Pc}}\xspace}

\def\bquark    {{\ensuremath{\Pb}}\xspace}

\def\pion   {{\ensuremath{\Ppi}}\xspace}
\def\piz    {{\ensuremath{\pion^0}}\xspace}

\def\pip    {{\ensuremath{\pion^+}}\xspace}
\def\pim    {{\ensuremath{\pion^-}}\xspace}

\def\rhomeson {{\ensuremath{\Prho}}\xspace}
\def\rhoz     {{\ensuremath{\rhomeson^0}}\xspace}

\def\kaon    {{\ensuremath{\PK}}\xspace}
  \def\Kbar    {{\kern 0.2em\overline{\kern -0.2em \PK}{}}\xspace}

\def\KorKbar    {\kern 0.18em\optbar{\kern -0.18em K}{}\xspace}

\def\Kp      {{\ensuremath{\kaon^+}}\xspace}
\def\Km      {{\ensuremath{\kaon^-}}\xspace}

\def\Kstarzb {{\ensuremath{\Kbar{}^{*0}}}\xspace}

\newcommand{\phiz}{\ensuremath{\Pphi}\xspace}
\newcommand{\omegaz}{\ensuremath{\Pomega}\xspace}

  \def\Dbar    {{\kern 0.2em\overline{\kern -0.2em \PD}{}}\xspace}
\def\D       {{\ensuremath{\PD}}\xspace}

\def\DorDbar    {\kern 0.18em\optbar{\kern -0.18em D}{}\xspace}
\def\Dz      {{\ensuremath{\D^0}}\xspace}

\def\Dp      {{\ensuremath{\D^+}}\xspace}

\def\Dstarp  {{\ensuremath{\D^{*+}}}\xspace}

\def\Ds      {{\ensuremath{\D^+_\squark}}\xspace}

\def\Bbar    {{\ensuremath{\kern 0.18em\overline{\kern -0.18em \PB}{}}}\xspace}

\def\BorBbar    {\kern 0.18em\optbar{\kern -0.18em B}{}\xspace}

\def\jpsi     {{\ensuremath{{\PJ\mskip -3mu/\mskip -2mu\Ppsi\mskip 2mu}}}\xspace}

  \def\Y#1S{\ensuremath{\PUpsilon{(#1S)}}\xspace}

\def\Lz          {{\ensuremath{\PLambda}}\xspace}
\def\Lbar        {{\ensuremath{\kern 0.1em\overline{\kern -0.1em\PLambda}}}\xspace}
\def\LorLbar    {\kern 0.18em\optbar{\kern -0.18em \PLambda}{}\xspace}

\def\Lc      {{\ensuremath{\Lz^+_\cquark}}\xspace}

\def\BF         {{\ensuremath{\cal B}}}

\newcommand{\decay}[2]{\ensuremath{#1\!\to #2}\xspace}         

\def\to                 {\ensuremath{\rightarrow}\xspace}

\newcommand{\msig}{{\ensuremath{m(\Km\pip\mup\mun)}}\xspace}
\newcommand{\mkppp}{{\ensuremath{m(\Km\pip\pip\pim)}}\xspace}

\def\signal    {\decay{\Dz}{\Km\pip\mup\mun}}
\def\norm    {\decay{\Dz}{\Km\pip\pip\pim}}

\def\AmpsignalT    {\Km\pip(\mup\mun)_{\rhoz-\omega}}

\def\Ampnorm    {\Km\pip\pip\pim}

\def\AT#1     {\ensuremath{A_{\mathrm{T}}^{#1}}\xspace}           

\def\C#1      {\ensuremath{\mathcal{C}_{#1}}\xspace}                       
\def\Cp#1     {\ensuremath{\mathcal{C}_{#1}^{'}}\xspace}                    
\def\Ceff#1   {\ensuremath{\mathcal{C}_{#1}^{\mathrm{(eff)}}}\xspace}        
\def\Cpeff#1  {\ensuremath{\mathcal{C}_{#1}^{'\mathrm{(eff)}}}\xspace}       
\def\Ope#1    {\ensuremath{\mathcal{O}_{#1}}\xspace}                       
\def\Opep#1   {\ensuremath{\mathcal{O}_{#1}^{'}}\xspace}                    

\newcommand{\tev}{\ifthenelse{\boolean{inbibliography}}{\ensuremath{~T\kern -0.05em eV}\xspace}{\ensuremath{\mathrm{\,Te\kern -0.1em V}}}\xspace}
\newcommand{\gev}{\ensuremath{\mathrm{\,Ge\kern -0.1em V}}\xspace}
\newcommand{\mev}{\ensuremath{\mathrm{\,Me\kern -0.1em V}}\xspace}
\newcommand{\kev}{\ensuremath{\mathrm{\,ke\kern -0.1em V}}\xspace}
\newcommand{\ev}{\ensuremath{\mathrm{\,e\kern -0.1em V}}\xspace}
\newcommand{\gevc}{\ensuremath{{\mathrm{\,Ge\kern -0.1em V\!/}c}}\xspace}
\newcommand{\mevc}{\ensuremath{{\mathrm{\,Me\kern -0.1em V\!/}c}}\xspace}
\newcommand{\gevcc}{\ensuremath{{\mathrm{\,Ge\kern -0.1em V\!/}c^2}}\xspace}
\newcommand{\gevgevcccc}{\ensuremath{{\mathrm{\,Ge\kern -0.1em V^2\!/}c^4}}\xspace}
\newcommand{\mevcc}{\ensuremath{{\mathrm{\,Me\kern -0.1em V\!/}c^2}}\xspace}

\def\mum  {\ensuremath{{\,\upmu\rm m}}\xspace}

\def\invpb {\ensuremath{\mbox{\,pb}^{-1}}\xspace}

\def\invfb   {\ensuremath{\mbox{\,fb}^{-1}}\xspace}

\newcommand{\chisq}{\ensuremath{\chi^2}\xspace}

\newcommand{\chisqip}{\ensuremath{\chi^2_{\rm IP}}\xspace}

\def\gsim{{~\raise.15em\hbox{$>$}\kern-.85em
          \lower.35em\hbox{$\sim$}~}\xspace}
\def\lsim{{~\raise.15em\hbox{$<$}\kern-.85em
          \lower.35em\hbox{$\sim$}~}\xspace}

\def\ptot       {\mbox{$p$}\xspace}
\def\pt         {\mbox{$p_{\rm T}$}\xspace}

\def\evtgen     {\mbox{\textsc{EvtGen}}\xspace}

\def\geant      {\mbox{\textsc{Geant4}}\xspace}

\def\photos     {\mbox{\textsc{Photos}}\xspace}

\def\pythia     {\mbox{\textsc{Pythia}}\xspace}

\def\tell1  {TELL1\xspace}
\def\ukl1   {UKL1\xspace}

\usepackage{cite} 
\usepackage{mciteplus}

\usepackage{longtable} 

\usepackage{mciteplus}
\usepackage{float}
\begin{document}

\renewcommand{\thefootnote}{\fnsymbol{footnote}}
\setcounter{footnote}{1}


\begin{titlepage}
\pagenumbering{roman}

\vspace*{-1.5cm}
\centerline{\large EUROPEAN ORGANIZATION FOR NUCLEAR RESEARCH (CERN)}
\vspace*{1.5cm}
\noindent
\begin{tabular*}{\linewidth}{lc@{\extracolsep{\fill}}r@{\extracolsep{0pt}}}
\ifthenelse{\boolean{pdflatex}}
{\vspace*{-2.7cm}\mbox{\!\!\!\includegraphics[width=.14\textwidth]{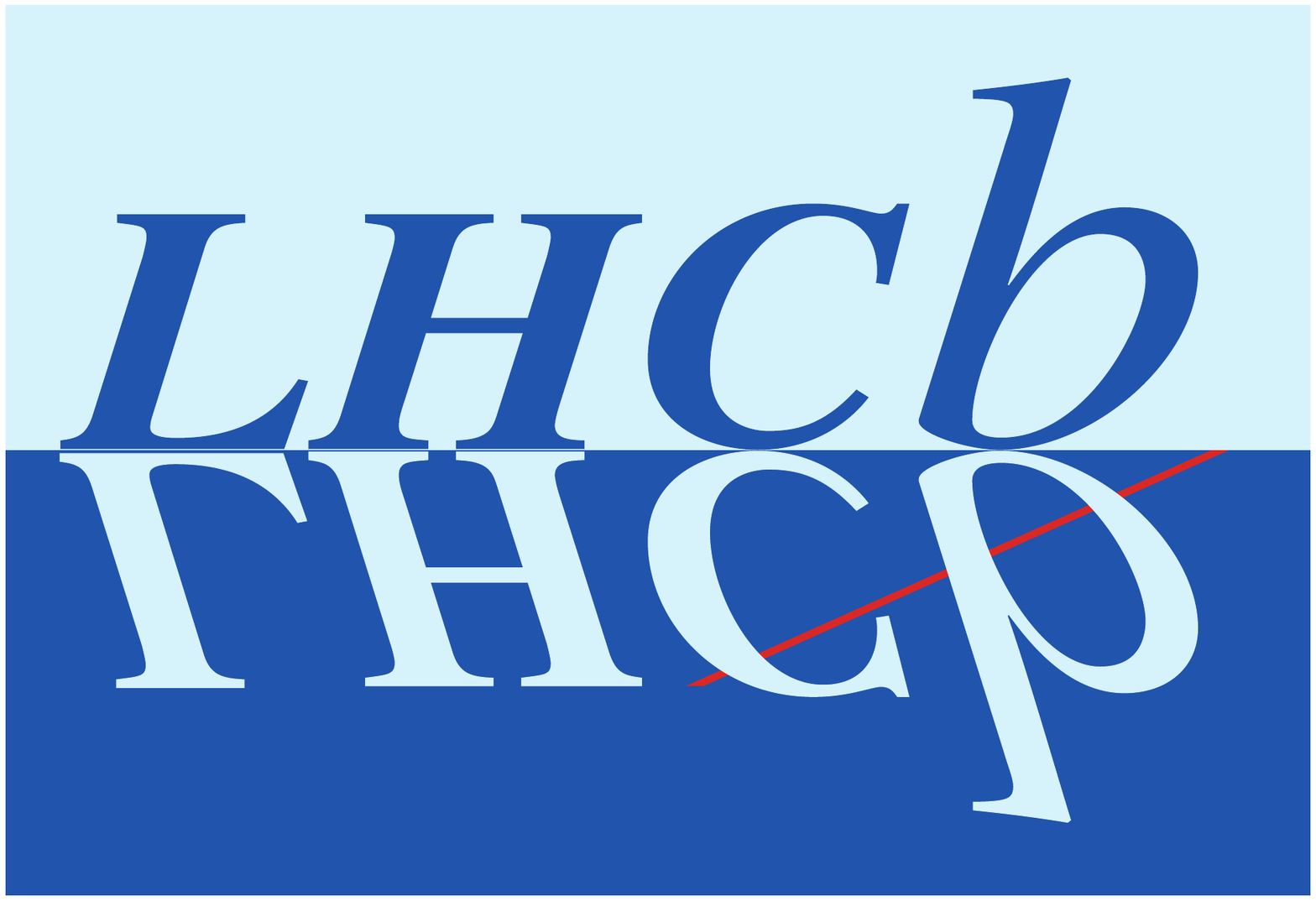}} & &}%
{\vspace*{-1.2cm}\mbox{\!\!\!\includegraphics[width=.12\textwidth]{lhcb-logo.eps}} & &}%
\\
 & & CERN-PH-EP-2015-283 \\  
 & & LHCb-PAPER-2015-043 \\  
 & & \today \\ 
 & & \\
\end{tabular*}

\vspace*{2.0cm}

{\bf\boldmath\huge
\begin{center}
First observation of the decay \signal in 
the \rhoz-\omegaz region of the dimuon mass spectrum
\end{center}
}

\vspace*{0.4cm}

\begin{center}
The LHCb collaboration\footnote{Authors are listed at the end of this Letter.}
\end{center}

\vspace*{.2cm}

\begin{abstract}
  \noindent
  A study of the decay \signal is performed using data collected by the LHCb detector in proton-proton collisions at a centre-of-mass energy of 8\tev, corresponding to an integrated luminosity of 2.0\invfb. Decay candidates with muon pairs that have an invariant mass in the range 675--875\mevcc are considered. This region is dominated by the $\rhoz$ and $\omega$ resonances. 
The branching fraction in this range is measured to be 
\begin{equation*}
\BF(\signal) = ( 4.17 \pm 0.12\,\mathrm{(stat)} \pm 0.40\,\mathrm{(syst)})\times10^{-6}. 
\end{equation*}
This is the first observation of the decay \signal. Its branching fraction is consistent with the value expected in the Standard Model.
\end{abstract}

\vspace*{1.0cm}

\begin{center}
  Published as Physics Letters B757 (2016) 558
\end{center}

\vspace{\fill}

{\footnotesize 
\centerline{\copyright~CERN on behalf of the \lhcb collaboration, licence \href{http://creativecommons.org/licenses/by/4.0/}{CC-BY-4.0}.}}
\vspace*{2mm}

\end{titlepage}


\newpage
\setcounter{page}{2}
\mbox{~}
%
%

\cleardoublepage


\renewcommand{\thefootnote}{\arabic{footnote}}
\setcounter{footnote}{0}



\pagestyle{plain} 
\setcounter{page}{1}
\pagenumbering{arabic}



\section{Introduction}
\label{sec:Introduction}

Rare charm decays may proceed via the highly suppressed {\decay{\cquark}{\uquark \mup\mun}} flavour
changing neutral current process. In the Standard Model such processes
can only occur through loop diagrams, where in charm decays the GIM
cancellation~\cite{GIM} is almost complete. As a consequence, the short-distance
contribution to the inclusive {\decay{\D}{X \mup\mun}} branching fraction is predicted
to be as low as {\ensuremath{\mathcal{O}(10^{-9})}}~\cite{PaulBigi:2011}, making these decays interesting for
searches for new physics beyond the Standard Model. However, taking into account
long-distance contributions through tree diagrams involving resonances
such as {\decay{\D}{X V(\rightarrow \mup\mun) }}, where $V$ represents a \phiz, \rhoz or \omegaz vector meson, 
the total branching fraction of these rare charm decays
can reach {\ensuremath{\mathcal{O}(10^{-6})}}~\cite{Fajfer:2007,PaulBigi:2011,Cappiello}. 
Their sensitivity to new physics therefore is greatest in regions of the dimuon mass spectrum away from these
resonances, where the main contributions to the branching fraction may come from short-distance amplitudes. 
Angular asymmetries are sensitive to new physics both in the vicinity of these resonances and away from 
them~\cite{Paul:2012ab,Fajfer:2015mia,deBoer:2015boa,Cappiello,Fajfer:2005ke} and could be as large as {\ensuremath{\mathcal{O}(1\%)}}.

This Letter focuses on the measurement of the decay\footnote{ The inclusion of charge conjugate decays is implied.} \signal. 
This will provide an important reference channel for measurements of the {\decay{\cquark}{\uquark \mup\mun}}
processes  \mbox{{\decay{\Dz}{\pi^{+}\pi^{-}\mup\mun}}} and {\decay{\Dz}{K^{+}K^{-}\mup\mun}}: precise branching fractions are easier
to obtain if they are compared with a normalisation mode that has similar features. When restricted to the dimuon mass range
\mbox{675 $ < m(\mup\mun) < $ 875\mevcc}, where the $\rhoz$ and $\omega$  resonances are expected to dominate, it can also 
be used to normalise the decays {\decay{\Dz}{K^{-}\pi^{+}\eta^{(')}(\rightarrow\mup\mun)}}.
Measuring their branching fractions allows the coupling $\eta^{(')}\rightarrow \mu^{+}\mu^{-}$ to be determined.
This contains crucial information for various low energy phenomena, and is an input to the prediction of the anomalous magnetic moment 
of the muon~\cite{KouEtaEtap,Masjuan:2015cjl,Nyffeler:2016gnb}. Focussing on this dimuon mass range also simplifies the analysis, which does not
have to account for the variation of the selection efficiency as a function of $ m(\mup\mun) $.
From previous measurements the most stringent 90\% confidence level upper limits on
the decay \mbox{\signal} are set by the E791 experiment~\cite{Aitala:2000kk}: \mbox{\BF(\signal) $<$  35.9~$\times$ 10$^{-5}$} in the full \Km\pip mass region and 
\BF(\signal) $<$  2.4~$\times$ 10$^{-5}$ in the region of the \Kstarzb resonance.

The study presented here is based on data collected by the LHCb detector in proton-proton collisions at a centre-of-mass energy of 8\tev, 
corresponding to an integrated luminosity of 2.0\invfb. A subsample corresponding to an integrated luminosity of 1.6\invfb
has been used to measure \BF(\signal). The remainder of the data set was used to optimise the selection. 
The branching fraction \BF(\signal) is measured relative to that of the normalisation decay
\norm. The most accurate recent measurement of this branching fraction is used, 
\BF(\norm)\,=\,(8.287\,$\pm$\,0.043\,$\pm$\,0.200)\,$\times$\,10$^{-2}$, 
obtained by the CLEO experiment~\cite{Bonvicini:2013vxi}.

\section{Detector and simulation}
\label{sec:DetAndSim}

The \lhcb detector~\cite{Alves:2008zz,LHCb-DP-2014-002} is a single-arm forward
spectrometer covering the \mbox{pseudorapidity} range $2<\eta <5$,
designed for the study of particles containing \bquark or \cquark
quarks. The detector includes a high-precision tracking system
consisting of a silicon-strip vertex detector surrounding the $pp$
interaction region, a large-area silicon-strip detector located
upstream of a dipole magnet with a bending power of about
$4{\mathrm{\,Tm}}$, and three stations of silicon-strip detectors and straw
drift tubes placed downstream of the magnet.
The tracking system provides a measurement of momentum, \ptot, of charged particles with
a relative uncertainty that varies from 0.5\% at low momentum to 1.0\% at 200\gevc.
The minimum distance of a track to a primary vertex, the impact parameter (IP), is measured with a resolution of $(15+29/\pt)\mum$,
where \pt is the component of the momentum transverse to the beam, in\,\gevc.

Different types of charged hadrons are distinguished using information
from two ring-imaging Cherenkov detectors.
Photons, electrons and hadrons are identified by a calorimeter system consisting of
scintillating-pad and preshower detectors, an electromagnetic
calorimeter and a hadronic calorimeter. Muons are identified by a
system composed of alternating layers of iron and multiwire
proportional chambers~\cite{LHCb-DP-2012-002}.

The online event selection is performed by a trigger~\cite{LHCb-DP-2012-004},
which consists of a hardware stage, based on information from the calorimeter and muon
systems, followed by a software stage, which applies a full event
reconstruction.
In the offline selection, requirements are made on whether the trigger decision was due to the signal 
candidate or to other particles produced in the $pp$ collision. Throughout this Letter, these two non-exclusive categories of candidates are referred to as
Trigger On Signal (TOS) and Trigger Independent of Signal (TIS) candidates. 

Simulated samples of \signal and \norm decays have been produced. 
In the simulation, $pp$ collisions are generated using
\pythia~\cite{Sjostrand:2006za,*Sjostrand:2007gs}
with a specific \lhcb configuration~\cite{LHCb-PROC-2010-056}.  Decays of hadronic particles
are described by \evtgen~\cite{Lange:2001uf}, in which final-state
radiation is generated using \photos~\cite{Golonka:2005pn}. The
interaction of the generated particles with the detector, and its response,
are implemented using the \geant toolkit~\cite{Allison:2006ve, *Agostinelli:2002hh} 
as described in Ref.~\cite{LHCb-PROC-2011-006}. No theoretical model or experimental measurement provides
a reliable decay model for \signal. This decay mode is therefore modelled as an incoherent sum of 
resonant and non-resonant contributions, such as \mbox{{\decay{\Kstarzb}{\Km\pip}}} and \mbox{{\decay{\rhoz$/$\omega}{\mup\mun}}}, 
motivated by the resonant structure observed in  \mbox{{\decay{\Dz}{\Km\pip\pip\pim}}} and  \mbox{{\decay{\Dz}{\Km\pip\pip\pim\piz}}} 
decays~\cite{PDG2014}, and by the theoretical predictions of Ref.~\cite{Cappiello}. In the case of \norm, a decay model reproducing 
the data was implemented using the MINT software package~\cite{Nason:2007vt}.

\section{Event selection}
\label{sec:Selection}

The criteria used to select the \signal and \norm decays are as similar as possible to allow 
many systematic uncertainties to cancel in the efficiency ratio.
At trigger level, only events that are TIS with respect to the hadron hardware trigger, 
which has a transverse energy threshold of 3.7\gev, are kept.
In the offline selection, the only differences between the signal and normalisation channels are the muon identification criteria. 

The first-level software trigger selects events that contain at least one good quality track
with high \pt and \chisqip, where the latter is defined as the difference in \chisq of the closest primary $pp$ interaction 
vertex (PV) reconstructed with and without the particle under consideration. The offline selection requires that at least 
one of these tracks originates from either the \signal or the \norm decay candidates.  
The second-level software trigger uses two dedicated selections to reconstruct
 \signal or \norm candidates originating from the PV. 
These combine good quality tracks that satisfy \mbox{\pt$>$ 350\mevc} and $p>$ 3000\mevc.    
A muon (\signal) or charged hadron (\norm) pair is required to form a good quality secondary vertex that is significantly
displaced from the PV. In events where such a pair is found, two charged hadrons are subsequently added.
The resulting four-particle candidate must have a good quality vertex and its invariant mass must be consistent 
with the known \Dz mass~\cite{PDG2014}. The momentum vector of this \Dz candidate must be consistent with having originated from 
the PV. 

A preselection follows the trigger selections. Four charged particles are combined to form \Dz candidates.
Tracks that do not correspond to actual trajectories of charged particles are suppressed by using
a neural network optimisation procedure. 
To reject 
the combinatorial background involving tracks from the PV, only high-$p$ and high-\pt tracks that are significantly displaced from any PV are used. 
This background is further reduced by requiring that the four decay products of the \Dz meson form a good quality vertex that is significantly displaced from 
the PV and that $p_{\mathrm{T}}$(\Dz) $>$ 3000\mevc. These three criteria also reject candidates formed from partially reconstructed charm hadron decays, combined with either random tracks from the PV or with tracks from the decay of another 
charmed hadron in the same event. 
This type of background is further reduced by requiring the \Dz momentum vector is within 14~mrad of the vector that joins the PV with the \Dz decay vertex, ensuring that
the \Dz candidate originates from the PV. Finally, the invariant mass of the \Dz candidate, which is reconstructed with a resolution of about 7\mevcc,  is required to 
lie within 65\mevcc of the known \Dz mass. 
In the case of \signal, $m(\mup\mun)$ is restricted to the range 675--875\mevcc. The two backgrounds described above are referred to as 
the non-peaking background throughout this Letter. 

After the preselection, a multivariate selection based on a boosted decision tree (BDT)~\cite{Breiman,Roe} is used to further suppress the non-peaking background. 
The GradBoost algorithm is used~\cite{TMVA}. The BDT uses the following variables: the \pt and \chisqip of the 
final state particles; the \pt and \chisqip of the \Dz candidate as well as the \chisq per degree of freedom of its vertex fit; the significance of 
the distance between this vertex and the PV; the largest distance of closest approach between the tracks that form the \Dz candidate; 
the angle between the \Dz momentum vector and the vector that joins the PV with its decay vertex. The cut on the BDT response 
used in the selection discards more than 80\% of the non-peaking candidates and retains more than 80\% of the signal candidates that have 
passed the preselection.

Finally, the information from the RICH, the calorimeters 
and the muon systems are combined to assign probabilities for each decay product to be a pion, a kaon or a muon, as 
described in Ref.~\cite{LHCb-DP-2014-002}.
A loose requirement on the kaon identification probability rejects about 90\% of the backgrounds
that consist of \pip\pim\mup\mun or \pip\pim\pip\pim combinations while preserving 98\% of the signal candidates. In the case of \signal decays, 
the muon identification criteria have an efficiency of 90\% per signal muon and reduce the rate of misidentified 
pions by a factor of about 150. In the absence of muon identification, \norm decays with two  misidentified pions would outnumber 
signal decays by four orders of magnitude. After these particle identification requirements, 
this background is reduced to around 50\% of the signal yield and is dominated by 
decays involving two pion decays in flight ({\decay{\pip}{\mup\neum}}). It is referred to as the peaking background throughout this Letter.

In addition to \norm decays with two misidentified pions, backgrounds due to the decays of \Dp, \Ds, \Dstarp, \tauon, \Lc and $\Sigma_{c}^{0}$
are considered. These are studied using simulated events and found to be negligible.

The selection is optimised using data and simulated samples. The BDT is trained using simulated \signal events to model the signal. 
The sample used to represent the background consists of candidates with $m$(\Kp\pim\mup\mun) $>$ 1890\mevcc, drawn from 2\% of the 
total data sample. Candidates on the low-mass side of the signal peak are not used due to the presence there of peaking background decays, whose
features are very close to those of signal decays. Optimal selection criteria on the BDT response and muon identification are found using another independent data sample corresponding to 20\% of 
the total dataset. The fit described in Sect.~\ref{sec:fitter} is used to estimate the yields of 
\signal signal ($S$), peaking background ($B_{\mathrm{pk}}$) and non-peaking background ($B_{\mathrm{npk}}$) present in this sample in the region of the signal peak, 
defined as \mbox{1840 $<$ \msig$<$ 1890\mevcc}. The requirements on the muon identification and BDT 
response are chosen to maximise $S/\sqrt{S+B_{\mathrm{pk}}+B_{\mathrm{npk}}}$. 

The two samples described above consist of events chosen randomly from the 2012 data and are not used for the subsequent analysis. 
The remainder of the dataset (78\%), which corresponds to an integrated luminosity of 1.6\invfb, is used to measure \mbox{\BF(\signal)}. The final \signal sample obtained 
with this selection consists of 5411 candidates. In the case of \norm, the large value of \BF(\norm) allows us to use a small sample (3\invpb), drawn 
randomly from the total dataset. The final \norm sample consists of 121\,922 candidates.

\section{Determination of the {\bf\boldmath \signal and \mbox{\norm}} yields}
\label{sec:fitter}

A simultaneous binned maximum likelihood fit to the \msig and \mkppp
distributions is performed to measure \mbox{\BF(\signal)}.


In each sample, the probability density function (PDF) fitted to the signal peak is a Gaussian function with power law tails. 
It is defined in the following way:
\begin{equation*}
\label{DCBForm}
f(m;m_{D^{0}},\sigma,\alpha_{L},n_{L},\alpha_{R},n_{R} )=
\left\lbrace
\begin{array}{cccc}
\dfrac{ \left(\frac{n_{L}}{|\alpha_{L}|}\right)^{n_{L}}\times e^{-\frac{1}{2}\alpha^{2}_{L}} }{\left(\frac{n_{L}}{|\alpha_{L}|}-|\alpha_{L}| -\frac{m-m_{D^{0}}}{\sigma}   \right)^{n^{\vphantom{s}}_{L}} } & \mbox{if} & \frac{m-m_{D^{0}}}{\sigma} \leq   -|\alpha_{L}|,\\
& & &  \\
\dfrac{ \left(\frac{n_{R}}{|\alpha_{R}|}\right)^{n_{R}}\times e^{-\frac{1}{2}\alpha^{2}_{R}} } {  \left(\frac{n_{R}}{|\alpha_{R}|}-|\alpha_{R}| + \frac{m-m_{D^{0}}}{\sigma}   \right)^{n^{\vphantom{s} }_{R}}  } & \mbox{if} & \frac{m-m_{D^{0}}}{\sigma} \geq   |\alpha_{R}|,\\  
& & &  \\ 
exp\left( \frac{ -(m-m_{D^{0}})^{2} }{ 2\sigma^{2} }\right ) & & \mbox{otherwise},\\
\end{array}\right.
\end{equation*}
where $m_{D^{0}}$ and $\sigma$ are the mean and width of the peak, and $\alpha_L$, $n_L$, $\alpha_R$ and $n_R$ parameterise the left and right tails. 
This function was found to describe accurately the \msig and \mkppp distributions obtained with the simulation,
which exhibit non-Gaussian tails on both sides of the peaks. The tail on the left-hand side is dominated by
final-state radiation and interactions with matter, while the right-hand side tail is due to non-Gaussian effects in the
reconstruction.

The non-peaking background in the \norm sample is described by a first-order polynomial. In the case of 
\signal, a second-order polynomial is used. 

Three peaking backgrounds due to misidentified \norm decays are categorised by the presence of 
candidates involving misidentified pions 
that did not decay in flight before reaching the most downstream tracking stations, or candidates where one or two pions 
decayed upstream of these tracking stations. Candidates from the first category are 
described by a one-dimensional kernel density estimate~\cite{Cranmer:2000du}. This PDF is derived from the \msig distribution obtained using simulated \norm decays 
reconstructed under the \signal hypothesis. Candidates from the remaining two categories appear as tails on the lower-mass side of the \msig distribution and must be accounted for to avoid biases in the non-peaking background and in the signal yield measured by the fit.  
Due to the small number of such candidates in the simulated sample, simulated \norm candidates where no pion decays in flight are 
altered to reproduce the effect of such decays, and the corresponding \msig distribution is determined.
This is achieved by modifying the momentum vectors of either one or two of the pions present in the \norm final state according to the kinematics 
of {\decay{\pip}{\mup\neum}} decays. The \msig distributions obtained after this modification are converted into one-dimensional kernel density estimates.

The fit model involves 5 yields: the signal yield, $N_{\mathrm{sig}}$, the yield of normalisation decays, $N_{\norm}$,
the peaking and non-peaking background yields, $N_{\mathrm{pk}}$ and $N_{\mathrm{npk}}$,  and the yield of
background candidates in the \norm sample, $N_{\mathrm{npk}}^{K\pi\pi\pi}$. They are all free parameters in the fit. 
It also involves 15 parameters to define the shapes of the PDFs. The parameters describing the widths and upper-mass tails
are free parameters in the fit but are common between the PDFs for the \signal and \norm peaks. 
The lower-mass tail parameters are determined separately. Those used for \norm candidates are allowed to vary in the fit. 
This is not possible for \signal candidates because of the overlap between the signal and the \norm peaking background and 
therefore the parameters are fixed to the values obtained from the simulated sample. In total, there are 15 free parameters
in the fit.

\begin{table}[t]
\caption{
\small Summary of the results of the fit described in Sect.~\ref{sec:fitter}. The yields measured in the \signal sample and the correlations between them, the yields measured in the normalisation sample, the common width fitted to the \signal and \norm yields, and the relative uncertainty on \mbox{\BF(\signal)} are presented. Uncertainties on the fitted parameters are statistical. 
The variation of the uncertainty on \mbox{\BF(\signal)} when the background yields 
are fixed indicates to what extent it is enhanced by the need to separate contributions in overlap and 
which shapes present some similarities.}
\begin{center}
\def\arraystretch{1.8}
\begin{tabular}{lc}
    \hline
     Parameter                      &   Value         \\ 
    \hline
      $N_{\mathrm{sig}}$       &       \,2357\,$\pm$\,67     \\ 
      $N_{\mathrm{pk}}$    &           \,1047\,$\pm$\,84  \\ 
      $N_{\mathrm{npk}}$   &           \,2007\,$\pm$\,116\\ 
      $N_{\mathrm{\norm}}$   &         \,83\,575\,$\pm$\,334\\ 
      $N_{\mathrm{npk}}^{K\pi\pi\pi}$   &    \,38\,346\,$\pm$\,257\\ 
      $\sigma$   &    7.17\,$\pm$\,0.03\,\mevcc \\ 
      $C_{N_{\mathrm{pk}}, N_{\mathrm{npk}}}$   & -78\%          \\ 
      $C_{N_{\mathrm{sig}}, N_{\mathrm{pk}}}$   & 27\%          \\ 
      $C_{N_{\mathrm{sig}}, N_{\mathrm{npk}}}$   & -48\%           \\ 
      $\sigma_{\BF(\signal)}$   &        2.9\%   \\ 
      $\sigma_{\BF(\signal)}$, if $N_{\mathrm{pk}}$ fixed   & 2.8\%          \\ 
      $\sigma_{\BF(\signal)}$, if $N_{\mathrm{pk}}$ and $N_{\mathrm{npk}}$ fixed   & 2.4\%          \\ 
    \hline
\end{tabular}
\end{center}
\label{tab:fitresults}
\end{table}

The relative yields of the three peaking background categories described above are fixed to values obtained by a fit to a large control sample. 
It consists of \signal candidates that are in the TOS category with respect to the muon hardware trigger, in 
contrast to the signal and normalisation samples that are in the TIS category with respect to the hadron trigger. All of the other selection requirements are the same as those described in Sect.~\ref{sec:Selection}. This TOS signal control sample consists of 28\,835 candidates and contains 
approximately six times more \signal decays than the nominal TIS sample.

\begin{figure}[p!]
  \begin{center}
    \includegraphics[width=0.9\linewidth]{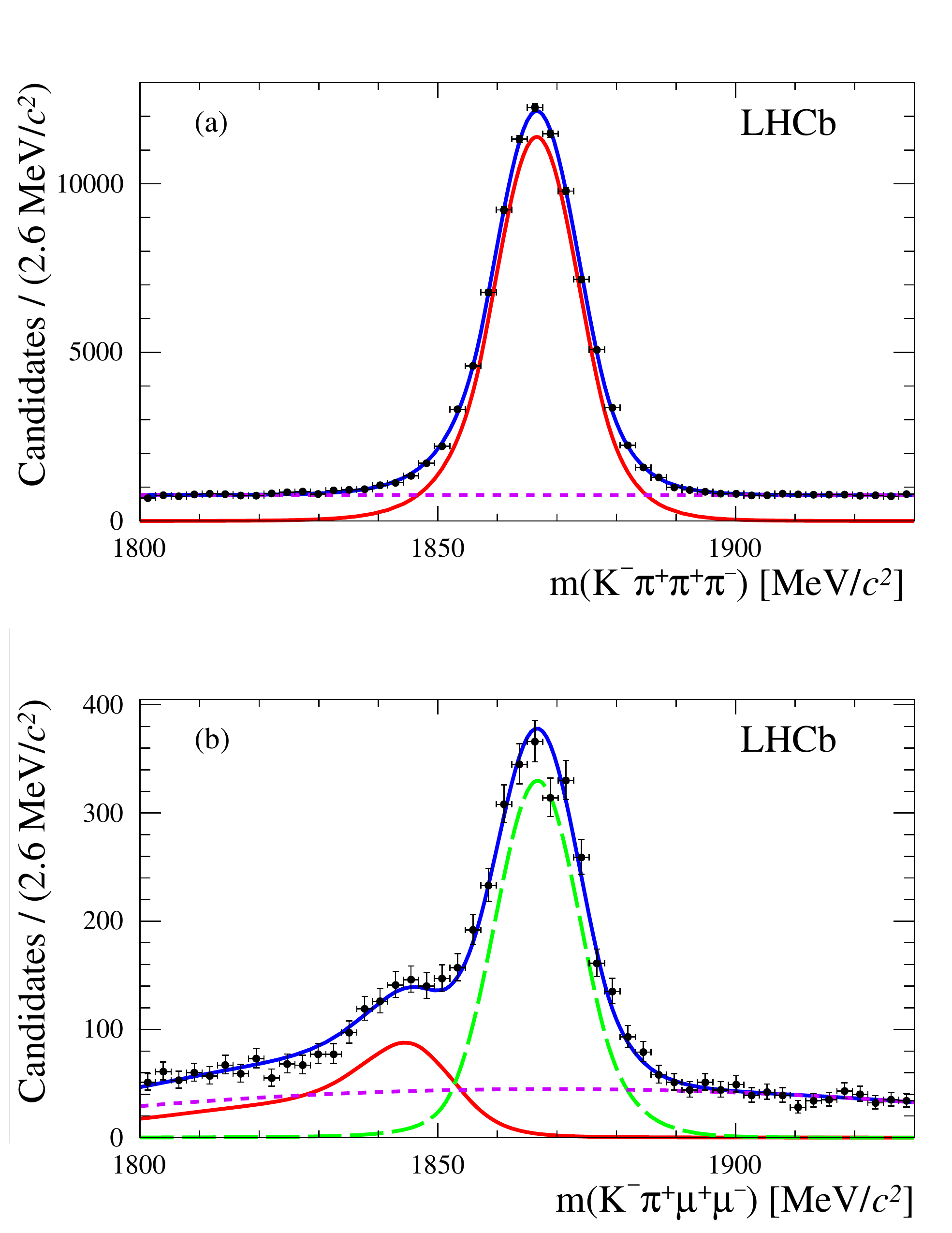}
    \vspace*{-0.3cm}
  \end{center}
  \caption{
    \small Mass distributions of  (a) \norm  and (b) \signal candidates. The data are shown as points (black) and the
total PDF (blue solid line) is overlaid. In (a), the two corresponding components of the fit model are the \norm decays (red solid line) and the non-peaking background (violet dashed line). 
In (b), the components are the \signal (long-dashed green line),  the peaking background due to misidentified \norm decays (red solid line),  and the non-peaking background (violet dashed line).} 
  \label{fig:fit}
\end{figure}

The fit results are summarized in Table ~\ref{tab:fitresults} and the observed mass distributions are shown in 
Fig.~\ref{fig:fit}, with fit projections overlaid. The main difficulties in this procedure are the similarities in the shape of the 
signal, peaking background and non-peaking background, and the overlap between their distributions in \msig. However, their
impact on the measurement presented in this Letter is limited, as can also be seen in Table ~\ref{tab:fitresults}.

\begin{figure}[p!]
  \begin{center}
    \includegraphics[width=0.8\linewidth]{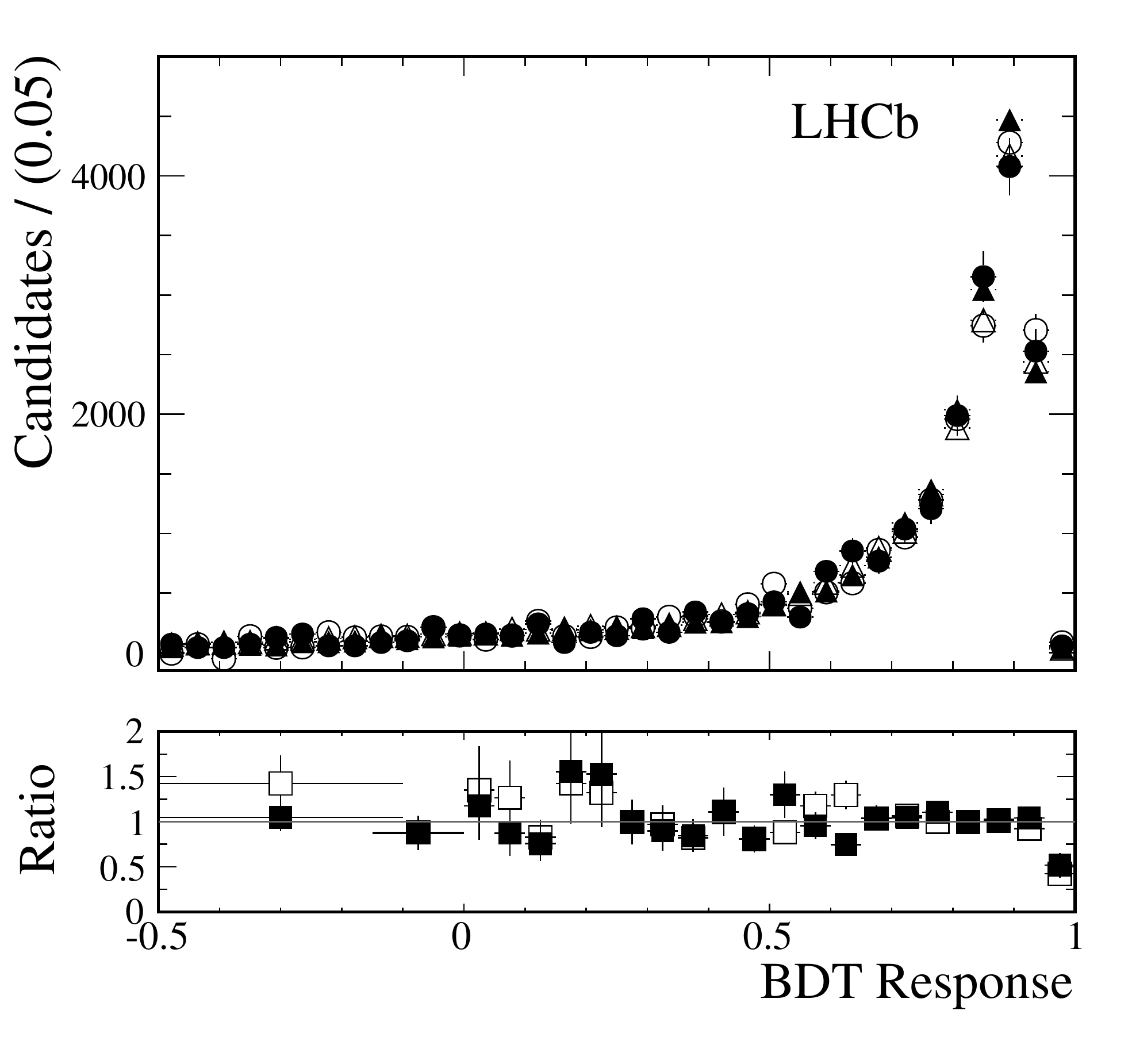}
    \vspace*{-0.5cm}
  \end{center}
  \caption{
    \small Distributions of the BDT response of \signal ( circles) and \norm decays (triangles) in data (full markers) and simulation 
(open markers). In data, the background contributions are removed using the {\it sPlot} technique.
The lower plot shows the ratio between the \signal and \norm distributions in data (full squares) and simulation
(open squares).}
  \label{fig:BDTfig}
\end{figure}

\section{Branching fraction measurement}
\label{sec:systematics}

The branching fraction of the decay \signal is obtained by combining the quantities presented in Table~\ref{tab:results}
with the branching fraction of the \norm decay according to
  \begin{equation}
    \label{eq:DeriveBF}
     \BF(\signal) = \frac{N_{\signal}}{N_{\norm}}\times\frac{\varepsilon_{\norm}}{\varepsilon_{\signal}}\times\BF(\norm),
\end{equation}
where $N_{\signal}$, $N_{\norm}$, $\varepsilon_{\signal}$ and $\varepsilon_{\norm}$ are the yields and selection efficiencies for the signal and 
normalisation decays. The branching fraction of the signal decay for dimuon invariant masses in the range 675--875\mevcc is measured to be
$\BF(\signal) = \mathrm{( 4.17 \pm 0.12)\times10^{-6}}$, where the uncertainty is statistical.

\begin{table}[t]
\caption{\small Measured efficiencies and yields for the decay \signal in the dimuon mass range 675--875\mevcc, and for the decay \norm.
The uncertainties are statistical. In the case of efficiencies, it stems from the finite size of the simulated samples.}
\begin{center}
\def\arraystretch{1.8}
\begin{tabular}{lcc}
    \hline
                        & {\footnotesize $\signal$} & {\footnotesize$\norm$}          \\ 
    \hline
    Efficiency [$\mathrm{10^{-5}}$]         & \ \ \ $8.8 \pm 0.2 $ & \ \ \ \ \ $8.2 \pm 0.1$   \\
    Yields              & $2357 \pm 67$ & \, $83\,575 \pm 334$  \\
    \hline
\end{tabular}
\end{center}
\label{tab:results}
\end{table}

\subsection{Systematic uncertainties}
\label{sec:realsystematics}

The systematic uncertainties on \BF(\signal) are summarised in Table~\ref{tab:systematics}. 
Those related to reconstruction and selection efficiencies are minimized thanks to the efficiency 
ratio in Eq.~\ref{eq:DeriveBF} and to the similarities between \signal and \norm decays. This is illustrated
in Fig.~\ref{fig:BDTfig}, which shows the distributions of the BDT response for the \signal and \norm decays, both 
in data and simulated samples. In data, the background contributions are removed using the {\it sPlot} technique~\cite{Pivk:2004ty}.
Also shown in this figure are the ratios between the  \signal and \norm distributions.
The BDT response, which combines all the offline selection variables (with the exception of muon identification criteria), 
is very similar for both kinds of decay and the differences are well described by the simulation. In cases where  
selection criteria depend on the nature of the decay products, data-driven methods are used, as described below.

The uncertainty on the charged hadron reconstruction inefficiency 
is dominated by the uncertainty on the probability to undergo a nuclear interaction in the detector.  This inefficiency is evaluated using simulated events. 
The corresponding uncertainty is derived from the 10\% uncertainty on the modelling of the detector material~\cite{LHCb-DP-2013-002}.

\begin{table}[t]
  \caption{\small Systematic uncertainties on \BF(\signal).}
\begin{center}
\def\arraystretch{1.}
\begin{tabular}{lc}
    \hline
      Source                 & Uncertainty [\%]         \\ 
    \hline
    \hline
      Track reconstruction                   &  3.2         \\ 
      Offline selection        &  2.0         \\ 
      Simulated decay models        &  2.5         \\ 
      Hardware trigger        &  4.4         \\ 
      Software trigger        &  4.3         \\ 
      Muon identification        &  3.2         \\ 
      Kaon identification        &  1.0         \\ 
      Size of simulated sample        &  2.9         \\ 
    \hline
    $\sigma_{\mathrm{syst}}(\varepsilon_{\signal}/\varepsilon_{\norm})$                  &  8.8         \\ 
    \hline
    \hline
      Signal shape parameters           &  0.8         \\ 
      Peaking background tails         &  1.5         \\ 
      Signal PDF                 & 0.6          \\ 
      Non-peaking background shape                 & 2.1          \\ 
    \hline
    $\sigma_{\mathrm{syst}}(N_{\AmpsignalT}/N_{\Ampnorm})$                  &  2.8         \\ 
    \hline
    \hline
     \BF(\norm)  & 2.5 \\
    \hline
    \hline
    Quadratic sum & 9.6 \\
    \hline
  \end{tabular}\end{center}
\label{tab:systematics}
\end{table}

The selection efficiencies based on the kinematical and geometrical requirements are derived from simulation. A systematic uncertainty 
to take into account imperfect track reconstruction modelling is 
estimated by smearing track properties to reproduce those observed in data. Similarly, a systematic uncertainty on the 
efficiency of the BDT selection is assigned as the difference between the efficiency obtained in data and simulation.

The uncertainties in the decay models are estimated separately for the signal and normalisation channels. For the signal, this is carried 
out by reweighting simulated \signal decays to reproduce the distributions of $m(\Km\pip)$ and $m(\mup\mun)$ observed in 
data, with the difference in 
efficiency relative to the default being assigned as the systematic uncertainty.
For \norm, the sensitivity to the decay model is studied by  comparing the default efficiency 
with that obtained in an extreme case in which the decay model provided by the MINT package is 
replaced by an incoherent sum of the resonances involved in the decay, as given in Ref.~\cite{PDG2014}.

To avoid dependence on the modelling of the hardware trigger in simulation, its efficiency is determined in data. 
The efficiency to be TIS with respect to hadron hardware trigger is determined as the fraction of 
\mbox{\signal} decays that fulfil this requirement among \mbox{\signal} candidates that are TOS with respect to the muon hardware trigger. 
It is measured in 12 different regions defined in the ($\pt(D^0)$, $N_{\mathrm{t}}$) plane, where $N_{\mathrm{t}}$ is the track multiplicity of the event. The overall hardware trigger efficiency for 
\mbox{\signal} decays is the average of these 12 efficiencies weighted according to the distributions of \mbox{\signal} candidates observed 
in data. The efficiency of the normalisation mode is obtained by weighting the same 12 efficiencies according to the distributions of 
\mbox{\norm} candidates.  This procedure assumes that the probability for \mbox{\signal} decays to fulfil the TIS 
requirement is not enhanced by the requirement to also be in the TOS category and that this TIS efficiency is the same in every region for \mbox{\signal} and 
\mbox{\norm} decays. 
No difference is found in simulation between 
the $\varepsilon_{\norm}/\varepsilon_{\signal}$ ratio obtained with this method and the ratio of true efficiencies, obtained by directly counting the
number of simulated \signal and \norm decays that fulfil the hadron trigger TIS requirement.
To determine the systematic uncertainty associated with the hardware trigger efficiency, the uncertainty on this comparison is combined with the statistical uncertainties on the 12 measurements performed in data in 
($\pt(D^0)$, $N_{\mathrm{t}}$) regions.

A similar approach is employed in the case of the first level of the software trigger. 
A sample of \norm candidates is selected from data that satisfied the trigger requirements
independently of these candidates. The fraction of \norm decays where at least one of the decay products 
also satisfies the requirements of this trigger is measured using this sample.
This efficiency is measured in regions of $\pt(\Dz)$ and weighted according to 
distributions of this variable in simulated \signal and \norm events. 
The variation in the efficiency ratio when these distributions are corrected to match the data is used to 
evaluate the corresponding systematic uncertainty. 

The efficiency of the second-level software trigger for the signal decay is calculated relative to that of the
normalisation decay. This ratio is measured using \signal decays in data and simulation and consistent results are obtained. 
The uncertainty on this comparison is therefore assigned as the systematic uncertainty
on this trigger efficiency. 

The efficiency of the muon identification criteria is determined in data using a large 
and pure sample of {\decay{B}{\jpsi (\rightarrow \mup\mun)X}} decays. Efficiencies measured 
in several regions of $p_{\mathrm{T}}(\mu)$, $\eta$($\mu$) and $N_{\mathrm{t}}$ are weighted according to the 
distribution observed for the muon candidates from \signal decays.  Several definitions of these domains are considered, with varying 
binnings. The different efficiencies obtained this way, as well as the efficiencies obtained in simulated samples,
are compared to evaluate the corresponding systematic uncertainty. The same approach is used to evaluate
the efficiency of the kaon identification requirement. In this case, the calibration kaons are provided
by {\decay{\Dstarp}{\Dz(\rightarrow\Km\pip)\pip}} decays in data.

In the fit outlined in Sect.~\ref{sec:fitter}, the parameters of the function 
that describe the lower-mass tail of the \signal peak are fixed to values obtained from simulation.
The corresponding systematic uncertainty is determined by repeating the fit using the values obtained by a
fit to the signal TOS control sample. A similar difference is observed when the corresponding test is performed for \norm 
candidates.

The systematic uncertainty related to the description of the peaking background 
is determined by the change observed in \BF(\signal) when the components due to the decay of one or two pions in flight are neglected,
and when their yields relative to the rest of the peaking background are enhanced by twice their uncertainty.

Two other systematic uncertainties have been evaluated. To estimate the impact of the signal PDF employed, 
the fit is repeated using the Cruijff function~\cite{delAmoSanchez:2010ae} instead. 
Potential effects arising from non-peaking backgrounds are assessed by repeating the fits
with the non-peaking backgrounds assumed to be linear in \msig. The values of the systematic uncertainties associated with 
the choice of fit model and its parameters were also further validated using pseudoexperiments. 

The impact on the fit of the similarities between the shapes of the signal and background
components was further controlled in two ways. First, fixing the background 
yields decreases the relative uncertainty on \BF(\signal) from 2.9\% to 2.4\%.
This variation is far lower than the total systematic uncertainty due to the 
yield determination (2.8\%). Moreover, another study is performed based on pseudoexperiments, 
generated with realistic values of the yields and PDFs shape parameters. The fit proved
able to return  unbiased measurements of the generated value of \BF(\signal) and an accurate
estimation of the statistical uncertainty, consistent with the uncertainty obtained 
in data.

As can be seen in Table~\ref{tab:systematics}, the systematic uncertainties are dominated by
the uncertainty on the \signal to \norm efficiency ratio, which is larger than the 2.9\%
statistical uncertainty on \BF(\signal). As expected, this systematic uncertainty
is primarily due to the different final state particles of the two decays.
The trigger efficiencies, and the muon identification and track reconstruction efficiencies, are 
responsible for about 90\% of this uncertainty. The uncertainties due to the 
yield determination and the knowledge of \BF(\norm) represent secondary 
contributions.

\begin{figure}[p!]
  \begin{center}
    \includegraphics[width=0.62\linewidth]{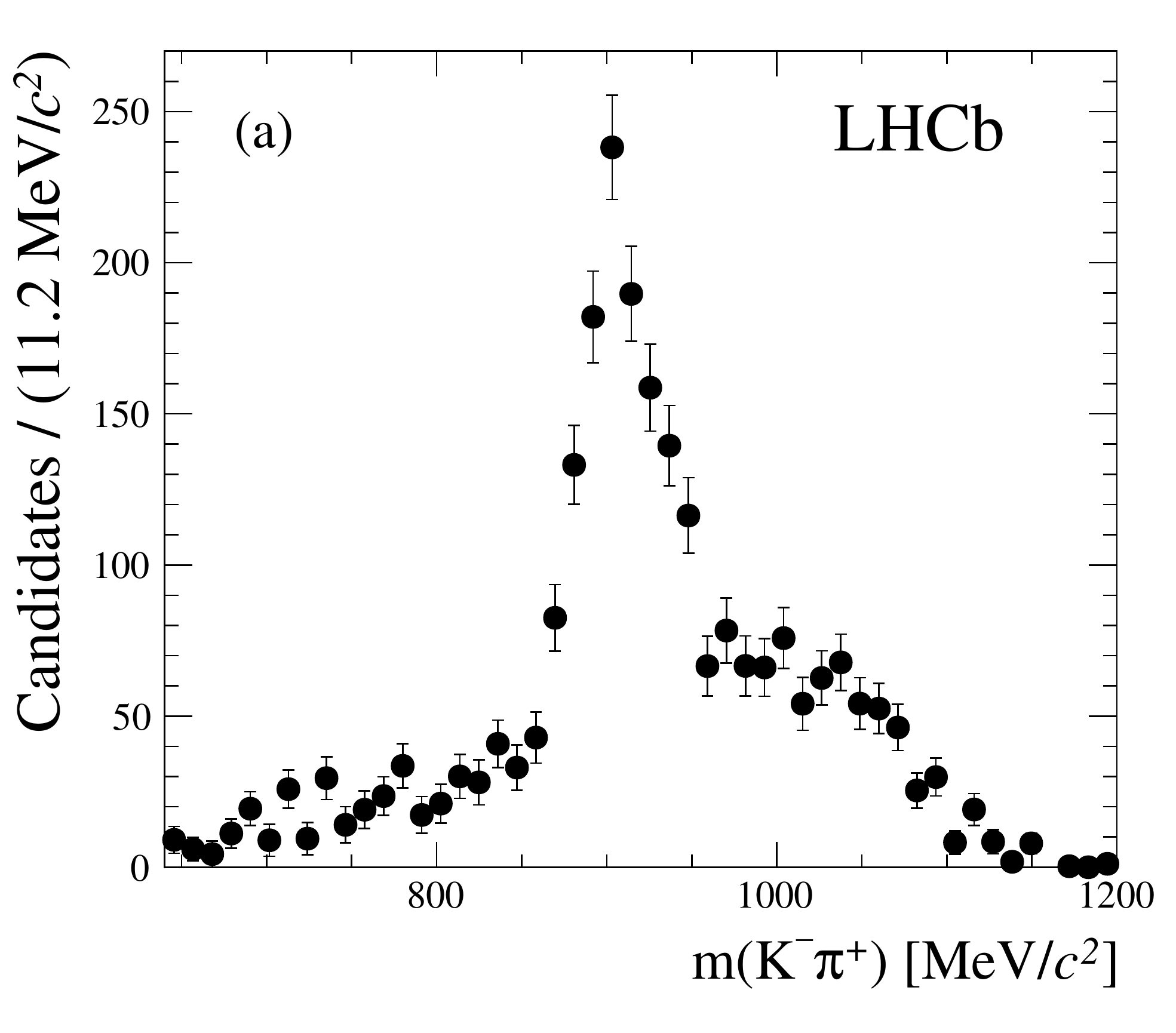}

     \includegraphics[width=0.61\linewidth]{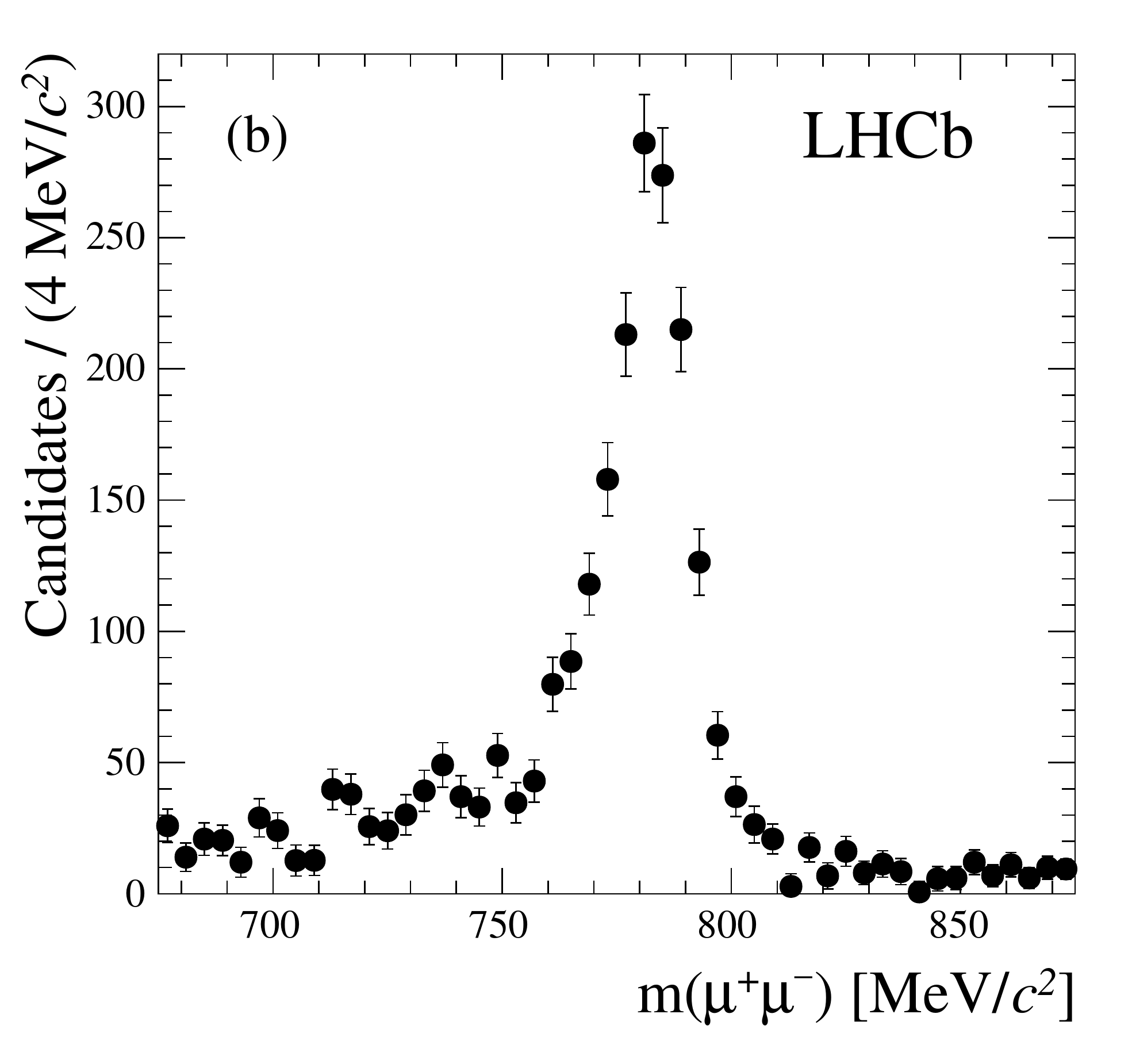}
    \vspace*{-0.5cm}
  \end{center}
  \caption{
    \small Background subtracted distribution of (a) the \Km\pip invariant mass and (b) the \mup\mun invariant mass,  measured in 
\signal decays using the {\it sPlot} technique.}
  \label{fig:dimuondihadron}
\end{figure}
\clearpage

\section{Conclusions}
\label{sec:conclu}

The decay \signal is studied using proton-proton collision data corresponding to an integrated luminosity 
of 2.0\invfb collected in 2012 by the LHCb detector at a centre-of-mass energy of 8\tev. 
The branching fraction of the decay \signal in the dimuon mass range 675--875\mevcc
is measured to be
  \begin{equation*}
    \BF(\signal) = ( 4.17 \pm 0.12\,\text{(stat)} \pm 0.40\,\text{(syst)})\times10^{-6}. 
  \end{equation*}

This branching fraction can be compared to the Standard Model value calculated 
in Ref.~\cite{Cappiello}, \BF(\signal)\,$=$\,6.7$\times$10$^{-6}$, in the full dimuon mass range.
This is the first observation of this decay. The branching fraction is measured with an overall precision of 10\% and is one order of
magnitude lower than the previous most stringent upper limit. Precise measurements of the {\decay{\Dz}{\pi^{+}\pi^{-}\mup\mun}} 
and {\decay{\Dz}{K^{+}K^{-}\mup\mun}} decays are now possible in all regions of the dimuon invariant 
mass since they can be compared with a normalisation mode that has similar features and a precisely known 
branching fraction. This will allow more stringent constraints on new physics to be obtained using data already collected by the LHCb detector, 
and the sensitivity of future experiments to angular asymmetries to be assessed.

The distributions of the \Km\pip and \mup\mun invariant masses in \signal decays 
are shown in Fig.~\ref{fig:dimuondihadron}\,, where the background contribution is 
removed using the {\it sPlot} technique~\cite{Pivk:2004ty}, taking the \msig invariant mass
as the discriminating variable. An amplitude analysis would be required for a full understanding of the decay dynamics.
The distributions in Fig.~\ref{fig:dimuondihadron}~~suggest the presence of additional contributions, including 
the \omegaz resonance, beyond the \Kstarzb\rhoz intermediate state that,  according to Ref.~\cite{Cappiello}, 
should strongly dominate the decay amplitude.

\section*{Acknowledgements}

\noindent We express our gratitude to our colleagues in the CERN
accelerator departments for the excellent performance of the LHC. We
thank the technical and administrative staff at the LHCb
institutes. We acknowledge support from CERN and from the national
agencies: CAPES, CNPq, FAPERJ and FINEP (Brazil); NSFC (China);
CNRS/IN2P3 (France); BMBF, DFG and MPG (Germany); INFN (Italy);
FOM and NWO (The Netherlands); MNiSW and NCN (Poland); MEN/IFA (Romania);
MinES and FANO (Russia); MinECo (Spain); SNSF and SER (Switzerland);
NASU (Ukraine); STFC (United Kingdom); NSF (USA).
We acknowledge the computing resources that are provided by CERN, IN2P3 (France), KIT and DESY (Germany), INFN (Italy), SURF (The Netherlands), PIC (Spain), 
GridPP (United Kingdom), RRCKI (Russia), CSCS (Switzerland), IFIN-HH (Romania), CBPF (Brazil), PL-GRID (Poland) and OSC (USA). We are indebted to the 
communities behind the multiple open source software packages on which we depend. We are also thankful for the
computing resources and the access to software R\&D tools provided by Yandex LLC (Russia).
Individual groups or members have received support from AvH Foundation (Germany),
EPLANET, Marie Sk\l{}odowska-Curie Actions and ERC (European Union),
Conseil G\'{e}n\'{e}ral de Haute-Savoie, Labex ENIGMASS and OCEVU,
R\'{e}gion Auvergne (France), RFBR (Russia), GVA, XuntaGal and GENCAT (Spain), The Royal Society
and Royal Commission for the Exhibition of 1851 (United Kingdom).

\bibliographystyle{LHCb}
\bibliography{main}
\newpage
\centerline{\large\bf LHCb collaboration}
\begin{flushleft}
\small
R.~Aaij$^{39}$, 
C.~Abell\'{a}n~Beteta$^{41}$, 
B.~Adeva$^{38}$, 
M.~Adinolfi$^{47}$, 
A.~Affolder$^{53}$, 
Z.~Ajaltouni$^{5}$, 
S.~Akar$^{6}$, 
J.~Albrecht$^{10}$, 
F.~Alessio$^{39}$, 
M.~Alexander$^{52}$
% ...

{\footnotesize \it
% ...
$ ^{5}$Clermont Universit\'{e}, Universit\'{e} Blaise Pascal, CNRS/IN2P3, LPC, Clermont-Ferrand, France\\
$ ^{6}$CPPM, Aix-Marseille Universit\'{e}, CNRS/IN2P3, Marseille, France\\
% ...
$ ^{10}$Fakult\"{a}t Physik, Technische Universit\"{a}t Dortmund, Dortmund, Germany\\
% ...
$ ^{38}$Universidad de Santiago de Compostela, Santiago de Compostela, Spain\\
$ ^{39}$European Organization for Nuclear Research (CERN), Geneva, Switzerland\\
% ...
$ ^{41}$Physik-Institut, Universit\"{a}t Z\"{u}rich, Z\"{u}rich, Switzerland\\
% ...
$ ^{47}$H.H. Wills Physics Laboratory, University of Bristol, Bristol, United Kingdom\\
% ...
$ ^{52}$School of Physics and Astronomy, University of Glasgow, Glasgow, United Kingdom\\
$ ^{53}$Oliver Lodge Laboratory, University of Liverpool, Liverpool, United Kingdom\\
% ...
}
\end{flushleft}
\newpage

\end{document}